\documentclass[aps,showpacs,prl,twocolumn,groupedaddress]{revtex4}
\usepackage{graphicx,epsfig}
\usepackage{amsmath}
\date{\today}
\usepackage[usenames]{color}
\usepackage{ulem} 

\newcommand{\bk}{{\bf k}}

\newcommand{\bt}{{\bf t}}

\newcommand{\bx}{{\bf x}}
\newcommand{\by}{{\bf y}}
\newcommand{\bz}{{\bf z}}
\newcommand{\bgamma}{\mbox{\boldmath$\gamma$}}

\newcommand{\bS}{{\bf S}}
\newcommand{\bT}{{\bf T}}

\newcommand{\kbt}{{k_{\rm B}T}}
\newcommand{\hot}{{\hat\bt}}

\begin{document}
\title{The anisotropic glassy properties of quasicrystals within an amended tunneling model}
%
%
\author{Drago\c s-Victor Anghel}
\affiliation{Department of Theoretical Physics, National Institute for Physics and Nuclear Engineering--''Horia Hulubei'', 30 Reactorului street, P.O.BOX MG-6, M\u agurele, Jud. Ilfov, Romania}
\author{Dmitry V. Churochkin}
\affiliation{Department of Physics, Faculty of Mathematical and Physical Sciences, University of Chile, Avenue Blanco Encalada 2008, Santiago, Chile}
%

\begin{abstract}
We use an extended version of the standard tunneling model to explain the anisotropic sound absorption in decagonal quasicrystals. The glassy properties are determined by an ensemble of two level systems (TLS), arbitrarily oriented. The TLS is characterized by a $3\times3$ symmetric tensor, which couples to the strain field through a $3\times3\times3\times3$ tensor of coupling constants, $[[R]]$. The structure of $[[R]]$ reflects the symmetry of the quasicrystal. 

We also analyze the probability distributions of the elements of $[T]$ in this particular model for a better understanding of the characteristics of ``isotropic'' and ``anisotropic'' orientations distributions of the ensemble of TLSs. We observe that the distribution of the elements is neither simple, nor intuitive and therefore it is difficult to guess it \textit{a priory}, using qualitative arguments based on the symmetry properties. 
\end{abstract}

\maketitle

\section{Introduction}\label{intro}

Despite of almost four decades of study of the glassy properties materials
\cite{PhysRevB.4.2029.1971.Zeller,Esquinazi:book,RevModPhys.74.991.2002.Pohl},
the nature of the two-level systems (TLS), the ubiquitous hypothetical
microscopic entities that are held responsible for these properties,
is still not known. As a general picture, it is accepted that they represent dynamic defects,
which are atoms or groups of atoms that tunnel from one minimum energy
configuration to another. But these atoms are in general not identified and,
even in the cases when they are identified (like in crystals with defects),
the TLS spectrum cannot be obtained based on a microscopic model.

The study of solids with anisotropic glassy properties is especially interesting since this brings additional information about the TLSs and requires a critical perspective on the standard tunneling model (STM).

To identify the origin of the TLS the  thermal and acoustical properties of a number of crystalline systems with defects \cite{SolidStateCommun19.335.1976.Lou,JPhysique.41.1193.1980.Doussineau,PhysRevB.42.5842.1990.Vanelstraete,PhysRevB.51.8158.1995.Laermans,PhysRevLett.81.3171.1998.Liu,Topp:thesis,PhysRevLett.75.1965.1995.Watson,ZPhysBCondensMatter.101.235.1996.Topp,PhysRevB.60.898.1999.Topp}
and quasicrystals \cite{PhysRevB.62.11437.2000.Thompson,PhysRevB.62.292.2000.Gianno,PhysRevLett.88.255901.2002.Bert}
at low temperatures have been investigated thoroughly. As expected,
the glass like properties in both the disordered crystalline system
and in the quasicrystals were revealed. However, a crucial difference
in comparison with ordinary amorphous solids which possess TLS
excitations  was detected. Namely, a pronounced anisotropy in internal
friction was clearly marked. As a consequence of that, the baffling
physical problem about the possible origin of the anisotropy
appeared. Indeed, there are two competitive ways for the explanation.
{\it First}, the effect is explained by an anisotropy in the distribution over the ``orientations'' of the TLSs in the ensemble. The interaction between the TLS and the elastic field is described by the Hamiltonian
\begin{eqnarray}
  H_{I} &=& \frac{1}{2}\left(\begin{array}{cc}
  \delta & 0 \\ 0 & -\delta
  \end{array}\right) , \label{HTLS_HI}
\end{eqnarray}
where $\delta\equiv 2\gamma_{ij}S_{ij}$ and $[S]$ is the strain field of the phonon--we assume everywhere summation over the repeated indices. The symmetric second rank tensor $[\gamma]$ characterizes the TLS and its ``deformability'' under elastic strain. 
The anisotropy of the physical properties is a reflection of the values taken by the elements of $[\gamma]$, which are determined by the lattice symmetries. 
%
%
Bert, Bellessa and Grushko made a conjecture regarding these values from which they eventually recovered the anisotropic sound attenuation rates in decagonal quasicrystals \cite{PhysRevLett.88.255901.2002.Bert}.


In the {\it second} approach \cite{PhysRevB75.064202.2007.Anghel,PhysRevB76.165425.2007.Kuhn,JPhysConfSer.92.12133.2007.Anghel}--the one that we shall employ in this paper--the TLS is characterized by a $3\times3$ symmetric tensor $[T]$ and the coupling between $[T]$ and $[S]$ is made through a forth rank tensor of coupling constants denoted by $[[R]]$. Explicitly, $\gamma_{ij}\equiv R_{klij}T_{kl}$ and $\delta\equiv[T]^t:[[R]]:[S]$. The elements of $[T]$ are determined by a unit vector $\hot$, which is the direction of the TLS, wheres the structure of  $[[R]]$ is determined by the symmetries of the host material. In this model, even if the TLSs are isotropically oriented, the anisotropy of the system is imposed by the properties of the tensor of coupling constants.

We applied the second approach to the crystalline materials of different symmetries with embedded TLSs, assuming the isotropy of the orientations of the TLSs \cite{PRB.78.94202.2008.Anghel,JPhysConfSer.150.012002.2009.Anghel,EPL.83.56004.2008.Anghel,RomJPhys.55.903.2010.Anghel} and we calculated the attenuation of ultrasound waves of different
polarizations and propagating in different crystallographic directions. Not only that this model describes in a simple way the asymmetries of the glassy properties, but it also allows us to make predictions about the relative attenuation rates of sound propagating in different directions.

In this paper we apply the model of Ref. \cite{PhysRevB75.064202.2007.Anghel} to the attenuation of ultrasound waves in quasicrystals and we obtain the attenuation rates along different crystallographic directions expressed in terms of the components of the tensor of coupling constants. We apply our calculations to the experimental results of Ref. \cite{PhysRevLett.88.255901.2002.Bert}, which enable us to calculate the relations between some of the components of the coupling constants tensor.


In order to better understand the distribution of the TLSs in isotropic and anisotropic materials, we calculate the distribution of the elements of $[T]$ for an isotropic distribution of unit vectors $\hot$. This distribution is neither simple, nor trivial, and by this we draw attention to the fact that it is difficult to guess it \textit{a priori}, without a more detailed model.

\section{The description of the TLS}\label{TLSmodel}

Let us introduce the notations by presenting briefly the model. The Hamiltonian of the free TLS is
\begin{equation}
  \label{eqn_TLS_hamiltonian}
  H_\text{TLS} = \frac{\Delta}{2}\sigma_z -\frac{\Lambda}{2}\sigma_x
  \equiv \frac{1}{2}\left(\begin{array}{cc}
                     \Delta   & -\Lambda\\
                     -\Lambda & -\Delta \end{array}
                \right)
\end{equation}
where $\Delta$ is called the {\em asymmetry of the potential} and $\Lambda$ the {\em tunnel splitting}. The eigenvalues of $H_\text{TLS}$ are $\pm\epsilon/2$, where $\epsilon\equiv\sqrt{\Delta^2+\Lambda^2}$ is the excitation energy of this TLS. The groundstate will be denoted by $|\! \downarrow\rangle$ and the excited state by $|\! \uparrow\rangle$. The interaction Hamiltonian of the TLS with the strain field is given by (\ref{HTLS_HI})

The parameters $\Delta$ and $\Lambda$ are distributed with the probability $P(\Delta,\Lambda)=P_0/\Lambda$, where $P_0$ is a constant. If expressed in terms of $\epsilon$ and $u\equiv\Lambda/\epsilon$, the probability distribution becomes $P(\epsilon,u)=P_0/(u\sqrt{1-u^2})$.

As usual, we work in the \textit{abbreviated subscript notations} and write $[S]$ and $[\gamma]$ as six-dimensional vectors: $\bS=(S_{11},S_{22},S_{33},2S_{23},2S_{13},2S_{12})^t$ and $\bgamma=(\gamma_{11},\gamma_{22},\gamma_{33},\gamma_{23},\gamma_{13},\gamma_{12})^t$, where the superscript $t$ denotes the transpose of a matrix or a vector.

As stated before, $\gamma_{ij}=R_{klij}T_{kl}$. Assuming that the TLS is characterized by a direction in space, $\hot$, the tensor $\bT$ is formed of the components of $\hot$ \cite{PhysRevB75.064202.2007.Anghel}. In abbreviated subscript notations $\bT\equiv(t_x^2,t_y^2,t_x^2,2t_yt_z,2t_zt_x,2t_xt_y)^T$, $R_{ijkl}$ becomes $R_{IJ}$, and $\bgamma\equiv [R]^T\cdot\bT$ \cite{PhysRevB75.064202.2007.Anghel}.
The structure of $[R]$ is determined by the symmetries of the lattice, since $\delta$ is a scalar and should be invariant under coordinates transformations, whereas $[R]$ should be invariant under the symmetry transformations that leaves the lattice invariant \cite{PhysRevB75.064202.2007.Anghel}.

The absorption rate of a phonon--with wavenumber $\bk$ and polarization $\sigma$--by a TLS is \cite{PhysRevB75.064202.2007.Anghel,JPhysConfSer.92.12133.2007.Anghel,PRB.78.94202.2008.Anghel,EPL.83.56004.2008.Anghel}
\begin{equation}
  \Gamma_{\bk\sigma}(\hat\bt) = \frac{2\pi}{\hbar}\frac{\Lambda^2n_{\bk\sigma}}{\epsilon^2}|\bT^t\cdot[R]\cdot\bS_{\bk\sigma}|^2\delta(\epsilon-\hbar\omega). \label{eqn_Gamma_bar}
\end{equation}
The main characteristic of the TLS-phonon interaction is contained in the quantity $M_{\bk,\sigma}(\hat\bt)\equiv\bT^t\cdot[R]\cdot\bS_{\bk\sigma}$, which bears an intrinsic anisotropy through the matrix $[R]$, on which the symmetries of the lattice are imposed.

The average scattering rate of a phonon by the ensemble of TLSs is obtained by averaging $\Gamma_{\bk\sigma}(\hat\bt)$ over $\epsilon$, $\Lambda$, and $\hat\bt$. In this way we get the total phonon absorption rate,
\begin{equation}
  \tau^{-1}_{\bk\sigma} = \frac{2\pi}{\hbar} P_0\langle|M_{\bk\sigma}(\hat\bt)|^2\rangle n_{\bk\sigma} \tanh\left(\frac{\epsilon}{2\kbt}\right), \label{av_def}
\end{equation}
which may be put into the standard form,
\begin{equation}
  \tau_{\bk,\sigma}^{-1} = \frac{2\pi}{\hbar} P_0 \gamma^2_{\hat\bk,\sigma}N^2 k^2 n_{\bk\sigma} \tanh\left(\frac{\epsilon}{2\kbt}\right), \label{av_STM}
\end{equation}
where $N=\sqrt{\hbar/(2V\rho\omega)}$ is the normalization constant of the phonon mode and $\gamma_{\hat\bk,\sigma}\equiv\langle|M_{\bk\sigma}(\hat\bt)|^2\rangle^{1/2}/(Nk)$ is the (average) phonon-TLS coupling constant.

We have no model regarding the orientations of the TLSs, so we shall assume that $\hat\bt$ is isotropically oriented.

The structure of $[R]$ should be similar to that of the elastic stiffness constants, $[c]$, so for a decagonal quasicrystal should have the form \cite{PhysRevLett.88.255901.2002.Bert,Phys-Usp.48.411.2005.Chernikov}
\begin{eqnarray}
[R] = \left(\begin{array}{cccccc}
r_{11}&r_{12}&r_{13}&0&0&0\\
r_{12}&r_{11}&r_{13}&0&0&0\\
r_{13}&r_{13}&r_{33}&0&0&0\\
0&0&0&r_{44}&0&0\\
0&0&0&0&r_{44}&0\\
0&0&0&0&0&r_{66}
\end{array}\right), && \label{R_trigonal}
\end{eqnarray}
where $r_{66}=(r_{11}-r_{12})/2$--the axis $z$ is taken along the tenfold axis. We observe that the structure of both, $[R]$ and $[c]$, is similar in decagonal quasicrystals and in hexagonal lattices.

Like in hexagonal lattices, in the decagonal quasicrystals we can have pure longitudinal and transversal waves propagating in all the three directions, $x$, $y$, and $z$.

The coupling constants, $\gamma_{\hat\bk,\sigma}$ are similar to those calculated for hexagonal lattices in \cite{EPL.83.56004.2008.Anghel}:
\begin{subequations} \label{gammasH}
\begin{eqnarray}
  \gamma_{k\hat\bx,l}^2  &=& \frac{2(r_{11}^{2}+r_{12}^{2}+r_{13}^{2})+(r_{11}+r_{12}+r_{13})^2}{15}, \nonumber \\
  &=& \gamma_{k\hat\by,l}^2 \label{gammaXlH} \\
  \gamma_{k\hat\bz,l}^2 &=& \frac{8r_{13}^{2}+4r_{13}r_{33}+3r_{33}^{2}}{15} , \label{gammaZlH} \\
  \gamma_{k\hat\by,\hat\bx,t}^2 &=& \gamma_{k\hat\bx,\hat\by,t}^2 = \frac{(r_{11}-r_{12})^2}{15} = \frac{4r_{66}^2}{15}, \label{gammaYXtH} \\
  \gamma_{k\hat\bx,\hat\bz,t}^2 &=& \gamma_{k\hat\by,\hat\bz,t}^2 = \gamma_{k\hat\bz,\hat\bx,t}^2 = \gamma_{k\hat\bz,\hat\by,t}^2 = \frac{4r_{44}^2}{15}, \label{gammaXZtH}
\end{eqnarray}
\end{subequations}
where by $l$ and $t$ we refer to longitudinal and transversal polarizations, respectively.

While for the longitudinal waves the direction of polarization is obvious, for the transversal waves the direction of polarization is indicated by the second unit vector in the subscript of $\gamma$, in the Eqs. (\ref{gammaYXtH}) and (\ref{gammaXZtH}).

Due to the isotropy condition in the decagonal plane, the coupling constants of the phonons propagating in this plane are independent of the direction of propagation if they have similar polarization.

Comparing now Eqs. (\ref{gammaYXtH}) and (\ref{gammaXZtH}) with the results of Bert et al. \cite{PhysRevLett.88.255901.2002.Bert}, we obtain
\begin{equation}
  \frac{P\gamma_\parallel^2}{P\gamma_\perp^2} \equiv \frac{\gamma_{k\hat\by,\hat\bx,t}^2}{\gamma_{k\hat\bx,\hat\bz,t}^2} = \left(\frac{r_{11}-r_{12}}{2r_{44}}\right)^2 = \left(\frac{r_{66}}{r_{44}}\right)^2 \approx 4.2, \label{rap_gammas}
\end{equation}
where $P\gamma_\parallel^2$ and $P\gamma_\perp^2$ are obvious notations from Ref. \cite{PhysRevLett.88.255901.2002.Bert}.

To find the ranges of the coupling constants, we rewrite the Eqs. (\ref{gammaXlH}) and (\ref{gammaZlH}) as
\begin{subequations}\label{gamma_mins}
\begin{eqnarray}
  \gamma_{k\hat\bx,l}^2  &=& \gamma_{k\hat\by,l}^2 =  \frac{r_{13}^2}{80}\Bigg[\Bigg(4\frac{r_{11}}{r_{13}}+1\Bigg)^2 + \Bigg(4\frac{r_{12}}{r_{13}}+1\Bigg)^2 \nonumber \\
  && + \frac{2}{3}\Bigg(4\frac{r_{11}}{r_{13}}+1\Bigg)\Bigg(4\frac{r_{12}}{r_{13}}+1\Bigg) + \frac{40}{3} \Bigg], \label{gammaXlH_min} \\
  \gamma_{k\hat\bz,l}^2 &=& \frac{r_{13}^{2}}{45}\left[\left(3\frac{r_{33}}{r_{13}}+2\right)^2+20\right] , \label{gammaZlH_min}
\end{eqnarray}
\end{subequations}
Since the function $f(a,b) = a^2 + b^2 + 2ab/3$ satisfies $f(a,b)\ge f(0,0) = 0$ for any $a$ and $b$, we find from Eqs.~(\ref{gamma_mins}) that
\begin{equation}
 \gamma_{k\hat\bx,l}^2  = \gamma_{k\hat\by,l}^2 \ge \frac{r_{13}^2}{6} \quad {\rm and} \quad \gamma_{k\hat\bz,l}^2 \ge \frac{4r_{13}^2}{9}. \label{cond_gamma_min}
\end{equation}
The minimum values of the coupling constants, $\gamma_{k\hat\bx,l}^2$ and $\gamma_{k\hat\bz,l}^2$, for given $r_{13}$ (\ref{cond_gamma_min}), are obtained for $r_{11}=r_{12}=-r_{13}/4$ and $r_{33} = -2r_{33}/3$. The ratio between these minimum values is 
\begin{equation}
 (\gamma_{k\hat\bx,l}^2)_{\rm min}/(\gamma_{k\hat\by,l}^2)_{\rm min} = 3/8. \label{gamma_ratio}
\end{equation}
and is independent of $r_{11}$, $r_{12}$, $r_{13}$, and $r_{33}$. 

\section{The anisotropy of the glassy properties}

The anisotropies of the glassy properties of a disordered system are determined by the properties of the tensor $\bgamma$. Bert, Bellessa and Grushko made in Ref. \cite{PhysRevLett.88.255901.2002.Bert} a  conjecture regarding the values of $\bgamma_{ij}$, which, they considered, would lead to the observed anisotropy.

The distributions we obtained in this paper are different from those conjectured in Ref. \cite{PhysRevLett.88.255901.2002.Bert}. In our model the ``orientations'' of the TLSs are well defined by the unit vectors $\hot$ and the symmetries of the host material are incorporated into the coupling constant tensor, $[R]$. Therefore the anisotropic distribution of the elements of the tensors $\bT$ and $\bgamma$ are clearly defined. This is not the case in the STM. To understand the difficulties to define by general, qualitative arguments the ``anisotropy'' of $\bgamma$ in the STM, we calculate in this section the probability distribution of the elements of our tensor $\bT$, under the assumption that the unit vectors $\hot$ are isotropically oriented. We shall see that these elements do not have all the same distribution of probability, nor the distributions are constant for the ranges of these variables. 

We define $\hot$ by the Euler angles, $\theta$ and $\phi$: $t_x=\sin\theta\cos\phi$, $t_y=\sin\theta\sin\phi$ and $t_z=\cos\theta$. Keeping $t_z$ and $\phi$ as variables, we write $\bT^t=[(1-t_z^2)\cos^2\phi, (1-t_z^2) \sin^2\phi, t_z^2, 2t_z\sqrt{1-t_z^2} \sin\phi, 2t_z\sqrt{1-t_z^2} \cos\phi, (1-t_z^2) \sin(2\phi)]$.

The variables $t_z$ and $\phi$ are uniformly distributed in the intervals $[0,1]$ and $[0,2\pi]$, respectively.
%
%
We denote their probability distribution by $P_\hot(t_z,\phi)=1/(4\pi)$. Then the probability distribution in the variables $T_I$ ($I=1,\ldots,6$) and $\phi$ is
\begin{equation}
  P_{I,\phi}(T_I,\phi) = \frac{1}{4\pi}\left|\frac{\partial(t_z,\phi)}{\partial(T_I,\phi)}\right| = \frac{1}{4\pi}\left|\frac{\partial(T_I,\phi)}{\partial(t_z,\phi)}\right|^{-1} \label{defP_T1_phi}
\end{equation}

The simplest of all is eventually
\begin{equation}
  P_{3,\phi}(T_3,\phi) = \frac{1}{4\pi\sqrt{T_3}}, \label{P_T3_phi}
\end{equation}
which, if integrated over $\phi$, gives
\begin{equation}
  P_{3}(T_3) = \frac{1}{2\sqrt{T_3}}. \label{P_T3}
\end{equation}

For the components $T_1$ and $T_2$ we obtain the probability distributions
\begin{subequations}\label{P_T1_T2_phi}
\begin{eqnarray}
  P_{1,\phi}(T_1,\phi) &=& \frac{1}{4\pi}\frac{\theta(\cos^2\phi-T_1)}{\sqrt{1-\frac{T_1}{\cos^2\phi}}\cos^2\phi}, \label{P_T1_phi} \\
  P_{2,\phi}(T_2,\phi) &=& \frac{1}{4\pi}\frac{\theta(\sin^2\phi-T_2)}{\sqrt{1-\frac{T_2}{\sin^2\phi}}\sin^2\phi}, \label{P_T2_phi}
\end{eqnarray}
\end{subequations}
where $\theta(x)$ is the Heaviside step function. An extra factor of 2 appears in the Eqs. (\ref{P_T1_T2_phi}) because $T_1$ and $T_2$ are even functions of $t_z$.

For an isotropic distribution of $\hot$, the components $T_1$, $T_2$, and $T_3$ are equivalent and using Eqs. (\ref{P_T3}) and (\ref{P_T1_T2_phi}) we obtain the identity
\begin{equation}
  P_{1}(T) = P_{2}(T) = \frac{1}{\pi}\int\limits_{\sqrt{T}}^1\frac{dx}{x\sqrt{(x^2-T_1)(1-x^2)}} \equiv \frac{1}{2\sqrt{T}}. \label{P_T1_T2_T3}
\end{equation}

\begin{figure}[t]
  \centering
  \includegraphics[width=6cm,bb=20 118 575 673]{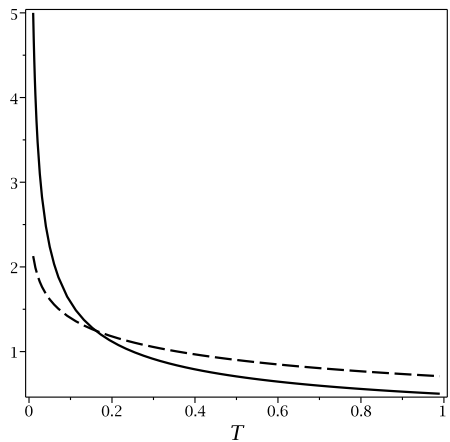}
  \caption{The probability distributions of the components of the tensor $\bT$: $P_{1}(T)=P_{2}(T)=P_{3}(T)=1/(2\sqrt{T})$, for $T\in(0,1]$ (solid line), and $2P_{4}(|T|)=2P_{5}(|T|)=2P_{6}(|T|)$, for $T\in[-1,0)\cup(0,1]$ (dashed line).}
  \label{P_T1_sqrt_T1}
\end{figure}

The components $T_4$, $T_5$, and $T_6$ are also equivalent for isotropic $\hot$. For example
\begin{subequations} \label{P_T6_T6_phi}
\begin{equation}
  P_{6,\phi}(T_6,\phi) = \frac{1}{4\pi}\frac{\theta(|\sin(2\phi)|-|T_6|)}{\sqrt{1-T_6/\sin(2\phi)}|\sin(2\phi)|}, \label{P_T6_phi}
\end{equation}
where $T_6/\sin(2\phi)=1-t_z^2\ge 0$. From Eq. (\ref{P_T6_phi}) we get
%
\begin{eqnarray}
  P_{6}(T) &=& \frac{1}{2\pi}\int_{|T|}^1\frac{dx}{\sqrt{x(x-T)(1-x^2)}} \label{P_T6} \\
  &=& P_{4}(T) =P_{5}(T), \label{P_T654}
\end{eqnarray}
\end{subequations}
where $T\in[-1,1]$ and $P_{4}(|T|)=P_{5}(|T|)=P_{6}(|T|)$.

We notice that $\lim_{T\to1}P_{T_6}(T)=1/(2\sqrt{2})$, whereas in 0, $P_{T_6}(T)$ has a logarithmic divergence.

The equalities (\ref{P_T654}) are obtained using the equivalence between the components $T_4$, $T_5$, and $T_6$.

We observe that the isotropic distribution of the TLS orientations do not correspond to a constant distribution of the values of the components of $\bT$, nor to equal distributions these values. Therefore it is not straightforward to draw conclusions regarding the relations between the distributions of the values of $\bgamma_I$, based only on general arguments. 

\section{Conclusions}

In this paper we describe the anisotropy of the glassy properties of the decagonal quasicrystals in the the model of Ref. \cite{PhysRevB75.064202.2007.Anghel}. We show that the glassy properties of these quasicrystals are similar to those of hexagonal disordered lattices \cite{EPL.83.56004.2008.Anghel} and we obtained the TLS-phonon average coupling constants, $\gamma^2_{\bk,\sigma}$, which are dependent on the phonon's propagation direction, $\bk$, and polarization, $\sigma$. We apply the results to the experimental data of Ref. \cite{PhysRevLett.88.255901.2002.Bert} and we obtain the ratio $(r_{66}/r_{44})^2 \approx 4.2$, where $r_{44}$ and $r_{66}$ are components of the tensor of coupling constants.

In order to better understand the characteristics of ``isotropic'' and ``anisotropic'' distributions of TLSs, we calculate the probability distributions of the elements of the tensor $\bT$ -- which describes the TLS -- under the assumption that the directions of the ensemble of TLSs, defined by the unit vectors $\hot$, are isotropically oriented. We observe that the distributions of the elements of $\bT$ are neither (all six) similar to each other, nor they are constant distributions. In conclusion, this distribution is not possible to find only by qualitative, general arguments, but one needs for this a deeper model, like the one of Ref. \cite{PhysRevB75.064202.2007.Anghel}.

\section*{Acknowledgements}
The work was supported by the Romanian National Authority for Scientific Research projects PN-II-ID-PCE-2011-3-0960 and PN09370102/2009. The travel support from the Romania-JINR Dubna collaboration project Titeica-Markov and project N4063 are gratefully acknowledged.


\end{document}